\documentclass{eptcs}
\providecommand{\event}{Axel Legay (Ed.): International Workshop on Verification of Infinite-State Systems (INFINITY)} 
\providecommand{\volume}{13}   
\providecommand{\anno}{2009}   
\providecommand{\firstpage}{1} 
\providecommand{\eid}{1}       
\usepackage{breakurl}        

\usepackage{amsmath}
\usepackage{amscd}
\usepackage{graphicx}
\usepackage{amssymb}
\usepackage{mathrsfs}
\usepackage{algorithmic}
\usepackage{algorithm}
\usepackage{amsmath,amssymb,alltt}
\newcommand{\suppress}[1]{} 

\newcommand{\QF}{F}

\newcommand{\St}{\mathit{St}}

\newcommand{\PP}{{\mathcal{P}}}

\newcommand{\EMPH}[1]{\emph{#1}}

\newtheorem{definition}{Definition}
\newtheorem{theorem}{Theorem}
\newtheorem{lemma}{Lemma}
\newtheorem{example}{Example}
\newtheorem{remark}{Remark}
\newtheorem{proof}{Proof}

\title{A Tighter Bound for the Determinization of Visibly Pushdown Automata\thanks{This research is partially supported by a COE-project}}
\author{Nguyen Van Tang
\institute{Research Center for Verification and Semantics\\
National Institute of Advanced Industrial Science and Technology\\
Toyonaka, Osaka, 560-0083 Japan} \email{t.nguyen@aist.go.jp}}

\def\titlerunning{A Tighter Bound for the Determinization of Visibly Pushdown Automata }
\def\authorrunning{N. V. Tang}

\begin{document}
\maketitle

\begin{abstract}
Visibly pushdown automata (VPA), introduced by Alur and Madhusuan in
2004, is a subclass of pushdown automata whose stack behavior is
completely determined by the input symbol according to a fixed
partition of the input alphabet. Since its introduce, VPAs have been
shown to be useful in various context, \emph{e.g.,} as specification
formalism for verification and as automaton model for processing XML
streams. Due to high complexity, however, implementation of formal
verification based on VPA framework is a challenge. In this paper we
consider the problem of implementing VPA-based model checking
algorithms. For doing so, we first present an improvement on upper
bound for determinization of VPA. Next, we propose simple on-the-fly
algorithms to check universality and inclusion problems of this
automata class. Then, we implement the proposed algorithms in a
prototype tool. Finally, we conduct experiments on randomly
generated VPAs. The experimental results show that the proposed
algorithms are considerably faster than the standard ones.
\end{abstract}

\section{Introduction}

Visibly pushdown automata~\cite{AM04} are pushdown automata whose
stack behavior (i.e. whether to execute a push, a pop, or no stack
operation) is completely determined by the input symbol according to
a fixed partition of the input alphabet. As shown in~\cite{AM04},
this class of visibly pushdown automata enjoys many good properties
similar to those of the class of finite automata. The main reason
for this being that is, each nondeterministic VPA can be transformed
into an equivalent deterministic one. Therefore, checking
context-free properties of pushdown models is feasible as long as
the calls and returns are made visible. As a result, visibly
pushdown automata have turned out to be useful in various context,
\emph{e.g.} as specification formalism for verification and
synthesis problem for pushdown systems~\cite{AEM04,ACM06,Loding}, as
automaton model for processing XML streams~\cite{Pitcher,KMV07}, and
as AOP protocols for component-based
systems~\cite{Nguyen06,Nguyen07}.

\medskip

As each nondeterministic VPA can be determinized, all problems that
concern the accepted languages such as universality and inclusion
problems are decidable. To check universality for a nondeterministic
VPA $M$ over its alphabet $\Sigma$ (that is, to check if
$L(M)=\Sigma^*$), the standard method is first to make it complete,
determinize it, complement it, and then checks for emptiness. To
check the inclusion problem $L(M)\subseteq L(N)$, the standard
method computes the complement of $N$, takes its intersection with
$M$ and then, check for emptiness. This is costly as computing the
complement necessitates a full determinization. This explosion is in
some sense unavoidable, because determinization for VPAs requires
exponential time blowup~\cite{AM04}. Therefore, one of the questions
raised is that whether one can implement efficiently operations like
determinization as well as decision procedures like universality (
or. inclusion) checking for VPAs.

\smallskip

During the recent years, a new approach called \textit{antichain
method} has been proposed to implement efficiently operations like
universality or inclusion checking on nondeterministic word or tree
automata~\cite{WDHR06,BHHTV08}. Unfortunately, the antichain
technique cannot be directly used for checking universality and
inclusion of VPA. This is because the set of configurations of a VPA
is infinite and thus, computing the set of antichains may not
terminate. In this paper, we focus on the problem of checking
universality and inclusion for VPAs. We make the following
contributions towards to this overall goal.

\begin{itemize}

\item First, we present an improvement on upper bound for
determinization of VPA. In~\cite{AM04}, Alur and Madhusudan showed
that any nondeterministic VPA with $n$ states can be translated into
a deterministic one with at most $2^{n^2 + n}$ states. Here, we show
that this upper bound can be made tighter. More precisely, we
optimize Alur-Madhusudan's determinization procedure, and show that
any nondeterministic VPA with $n$ states can be transformed into a
deterministic one with at most $2^{n^2}$ states.

\item Second, we apply the standard method to check universality and
inclusion problems for nondeterministic VPA. This method includes
two main steps: determinization and reachability checking for
non-accepting configurations. For determinization, we use the
Alur-Madhusudan's procedure~\cite{AM04}. For reachability checking,
we apply the symbolic technique
$\PP$-automata~\cite{Esparza00,Esparza03} to compute the sets of all
reachable configurations of a VPA.

\item Third, we present an on-the-fly method to check universality
of VPA. The idea is very simple that we perform determinization and
reachability checking by $\PP$-automaton simultaneously. For
checking universality of nondeterministic VPA $M$, we first create
the initial state of the determinized VPA $M^d$ and, initiate a
$\PP$-automaton $A$ to represent the initial configuration of $M^d$.
Second, construct new transitions departing from the initial states,
and update the $\PP$-automaton $A$. Then, the determinized VPA $M^d$
is updated using new states and transitions of $A$ (which correspond
to pairs of the states and topmost stack symbols of $M^d$), and so
on. When a non-accepting state is added to $A$, we stop and report
that $M$ is not universal.

\item Fourth, we also propose a new algorithmic solution to inclusion checking for
VPAs using on-the-fly manner. Again, no explicit determinization is
performed. To solve the language-inclusion problem for
nondeterministic VPAs, $L(M)\subseteq L(N)$, the main idea is to
find at least one word $w$ accepted by $M$ but not accepted by $N$,
\emph{i.e.,} $w\in L(M)\setminus L(N)$.

\item Finally, we have implemented all algorithms in
a prototype tool (written in Java 1.5) and tested them in a series
of experiments. Although the standard methods (as well as on-the-fly
ones) have the same worst case complexity, our preliminary
experiments on randomly generated visibly pushdown automata show a
significant improvement of on-the-fly methods compared to the
standard ones.

\end{itemize}

The remainder of this paper is organized as follows. In
Section~\ref{Sec:VPA} we recall notions and properties of VPAs, and
then we give an improvement on determinization of VPAs.
Section~\ref{Sec:Decision} presents new algorithms for checking
universality and inclusion of VPAs. Implementation as well as
experimental results are presented and analyzed in
Section~\ref{Sec:Implement}. Section~\ref{Sec:Related} discusses
about related works. Finally, we conclude the paper in
Section~\ref{Sec:Conclusion}.

\section{Visibly Pushdown Automata}\label{Sec:VPA}
\subsection{Definitions}

In this section we briefly recall the notions and properties of
visibly pushdown automata. Readers are referred to the seminal
paper~\cite{AM04} for their more details.

\medskip

Let $\Sigma$ be the finite input alphabet, and let $\Sigma =
\Sigma_c \cup \Sigma_r \cup \Sigma_i$ be a partition of $\Sigma$.
The intuition behind the partition is: $\Sigma_c$ is the finite set
of \EMPH{call} (push) symbols, $\Sigma_r$ is the finite set of
\EMPH{return} (pop) symbols, and $\Sigma_i$ is the finite set of
\EMPH{internal} symbols. Visibly pushdown automata are formally
defined as follows:

\begin{definition}\label{Def:VPA}
A \EMPH{visibly pushdown automaton} (VPA) $M$ over $\Sigma$ is a
tuple $(Q, Q_0, \Gamma, \Delta, \QF)$ where $Q$ is a finite set of
states, $Q_0 \subseteq Q$ is a set of initial states, $\QF \subseteq
Q$ is a set of final states, $\Gamma$ is a finite stack alphabet
with a special symbol $\bot$ (representing the
\EMPH{bottom-of-stack}), and $\Delta =
\Delta_c\cup\Delta_r\cup\Delta_i$ is the transition relation, where
$\Delta_c \subseteq Q\times\Sigma_c \times Q\times
(\Gamma\setminus\{\bot\})$, $\Delta_r \subseteq
Q\times\Sigma_r\times\Gamma\times Q$, and $\Delta_i \subseteq
Q\times\Sigma_i\times Q$.
\end{definition}

If $ (q, c, q',\gamma) \in\Delta_c$, where $c\in\Sigma_c$ and
$\gamma\neq\bot$, there is a \emph{push-transition} from $q$ on
input $c$ where on reading $c$, $\gamma$ is pushed onto the stack
and the control changes from state $q$ to $q'$; we denote such a
transition by $q \xrightarrow{c/+\gamma} q'$. Similarly, if $(q, r,
\gamma, q')$, there is a \emph{pop-transition} from $q$ on input $r$
where $\gamma$ is read from the top of the stack and popped (if the
top of the stack is $\bot$, then it is read but not popped), and the
control state changes from $q$ to $q'$; we denote such a transition
$q \xrightarrow{r/-\gamma} q'$. If $(q, i, q')\in\Delta_i$, there is
an \emph{internal-transition} from $q$ on input $i$ where on reading
$i$, the state changes from $q$ to $q'$; we denote such a transition
by $q \xrightarrow{i} q'$. Note that there are no stack operations
on internal transitions.

\smallskip

We write $\St$ for the set of \EMPH{stacks} $\{ w\bot \mid w \in
(\Gamma \setminus \{\bot\})^* \}$. A {\em configuration} is a pair
$(q,\sigma)$ of $q \in Q$ and $\sigma \in \St$. The transition
function of a VPA can be used to define how the configuration of the
machine changes in a single step: we say $(q,\sigma)\xrightarrow{a}
(q',\sigma')$ if one of the following conditions holds:
\begin{itemize}
\item If $a\in\Sigma_c$ then there exists $\gamma\in\Gamma$ such that $q \xrightarrow[]{a/+\gamma}
q'$ and $\sigma'=\gamma\cdot\sigma$
\item If $a\in\Sigma_r$, then there exists $\gamma\in\Gamma$ such that $q \xrightarrow{a/-\gamma} q'$
and either $\sigma = \gamma\cdot\sigma'$, or $\gamma=\bot$ and
$\sigma = \sigma' = \bot$
\item If $a\in\Sigma_i$, then $q \xrightarrow{a} q'$ and $\sigma=\sigma'$.
\end{itemize}

A $(q_0, w_0)$-{\em run} on a word $u = a_1\cdots a_n$ is a sequence
of configurations $(q_0, w_0) \mathrel{\buildrel a_1 \over
\rightarrow} (q_1, w_1) \cdots
        \mathrel{\buildrel a_n \over \rightarrow} (q_n, w_n)$, and
is denoted by $(q_0,w_0) \mathrel{\buildrel u \over \rightarrow}
(q_n, w_n)$. A word $u$ is accepted by $M$ if there is a run
$(q_0,w_0) \mathrel{\buildrel u \over \rightarrow} (q_n, w_n)$ with
$q_0 \in Q_{0}$, $w_0 = \perp$, and $q_n \in Q_F$. The language
$L(M)$ is the set of words accepted by $M$. The language $L
\subseteq \Sigma^*$ is a \EMPH{visibly pushdown language} (VPL) if
there exists a VPA $M$ with $L = L(M)$.

\medskip

\begin{definition}
{\it A VPA $M$ is {\em deterministic} if $|Q_{0}| = 1$ and for every
configuration $(q,\sigma)$ and $a \in \Sigma$, there are at most one
transition from $(q,\sigma)$ by $a$. For deterministic VPAs (DVPAs)
we denote the transition relation by $\delta$ instead of $\Delta$,
and write:

\begin{enumerate}
\item $\delta (q,a) = (q',\gamma)$ instead of $(q, a,
q',\gamma)\in\Delta$ if $a\in\Sigma_c$,

\item $\delta(q, a, \gamma)
= q'$ instead of $(q, a, \gamma, q')\in \Delta$ if $a\in\Sigma_r$,
and

\item  $\delta (q,a) = q'$ instead of $(q, a, q')\in\Delta$ if $a\in
\Sigma_i$.
\end{enumerate}}
\end{definition}

\subsection{Determinization}

As shown in~\cite{AM04}, any nondeterministic VPA can be transformed
into an equivalent deterministic one. The key idea of the
determinization procedure is to do subset construction, but
postponing handling push transitions. The push transitions are
stored into the stack and simulated at the time of matching pop
transitions. The construction has two components: a set of
\emph{summary edges} $S$, that keeps track of what state transitions
are possible from a push transition to the corresponding pop
transition, and a set of \emph{path edges} $R$, that keeps track of
all possible state reached from initial states. For completeness,
let us briefly recall the original determinization
procedure~\cite{AM04} as below.

\medskip

Let $M = (Q, \Gamma, Q_{0}, \Delta, F)$ be a nondeterministic VPA.
We construct an equivalent deterministic VPA $M' = (Q', \Gamma',
Q'_{0}, \Delta',F' )$ as follows: $Q'= 2^{Q\times Q}\times 2^{Q}$,
$Q'_{0}=\{(Id_Q, Q_0)\}$ where $Id_Q = \{(q,q) ~|~ q\in Q\}$, $F' =
\{(S,R) ~|~ R\cap F\neq\varnothing\}$, $\Gamma'= Q'\times\Sigma_c$,
and the transition relation
$\Delta'=\Delta'_i\cup\Delta'_c\cup\Delta'_r$ is given by:
\begin{itemize}
\item {\bf Internal:} For every $a\in\Sigma_i$,
$(S,R)\xrightarrow{a} (S',R')\in\Delta'_i$ where $ S' = \{(q,q') ~|~
\exists q^{\prime\prime}\in Q: (q,q^{\prime\prime})\in S,
q''\xrightarrow{a} q'\in \Delta_i\}$, and $R' =  \{q' ~|~ \exists
q\in R: q\xrightarrow{a} q'\in\Delta_i\}$.

\item {\bf Push:} For every $a\in\Sigma_c$,
$(S,R)\xrightarrow{a/+(S,R,a)} (Id_Q,R') \in\Delta'_c$ where $R' =
\{q' ~|~ \exists q\in R: q\xrightarrow{a/+\gamma} q'\in\Delta_c\}$.

\item {\bf Pop:} For every $a\in\Sigma_r$,
\begin{itemize}
\item if the stack is empty : $(S,R)\xrightarrow{a/-\perp}(S',R')\in\Delta'_r$ where
$S' = \{(q,q')~|~\exists q^{\prime\prime}\in Q:
    (q,q^{\prime\prime})\in S, q''\xrightarrow{a/-\perp}
    q'\in\Delta_r\}$ and $R' = \{q' ~|~ \exists q\in R: q\xrightarrow{a/-\perp}
    q'\in\Delta_r\}$.

\item otherwise:
$(S,R)\xrightarrow{a/-(S',R',a')}(S^{\prime\prime},R^{\prime\prime})\in\Delta'_r$,
where
$$\left\{
  \begin{array}{lll}
   R^{\prime\prime} & = &
    \left\{ q'
    \left|
    \begin{array}{l@{}l}
    \exists q\in R': (q,q')\in Update
    \end{array}
    \right.
    \right\} \vspace{1mm}\\

   S^{\prime\prime} & = &
   \{(q,q')~|~\exists q_3\in Q: (q,q_3)\in S', (q_3,q')\in Update\}
    \vspace{1mm}\\

   Update & = &
    \left\{  (q,q')
    \left|
    \begin{array}{l@{}l}
    \exists q_1\in Q, q_2\in R: (q_1,q_2)\in S, \\
   q\xrightarrow{a'/+\gamma}q_1\in \Delta_c,
   q_2\xrightarrow{a/-\gamma} q'\in \Delta_r \\
    \end{array}
    \right.
    \right\} \\

  \end{array}
  \right.$$
\end{itemize}
\end{itemize}

\begin{theorem}[\textnormal{\cite[Theorem~2]{AM04}}]
\label{th:VPAdeterminization} {\it Let $M$ be a VPA. The VPA $M'$ is
deterministic and $L(M') = L(M)$.  Moreover, if $M$ has $n$ states,
one can construct $M'$ with at most $2^{n^2+n}$ states and with
stack alphabet of size $|\Sigma_c|\cdot 2^{n^2+n}$.}
\end{theorem}

\begin{figure}
\begin{center}
  \includegraphics[height=3.5in,width=6in]{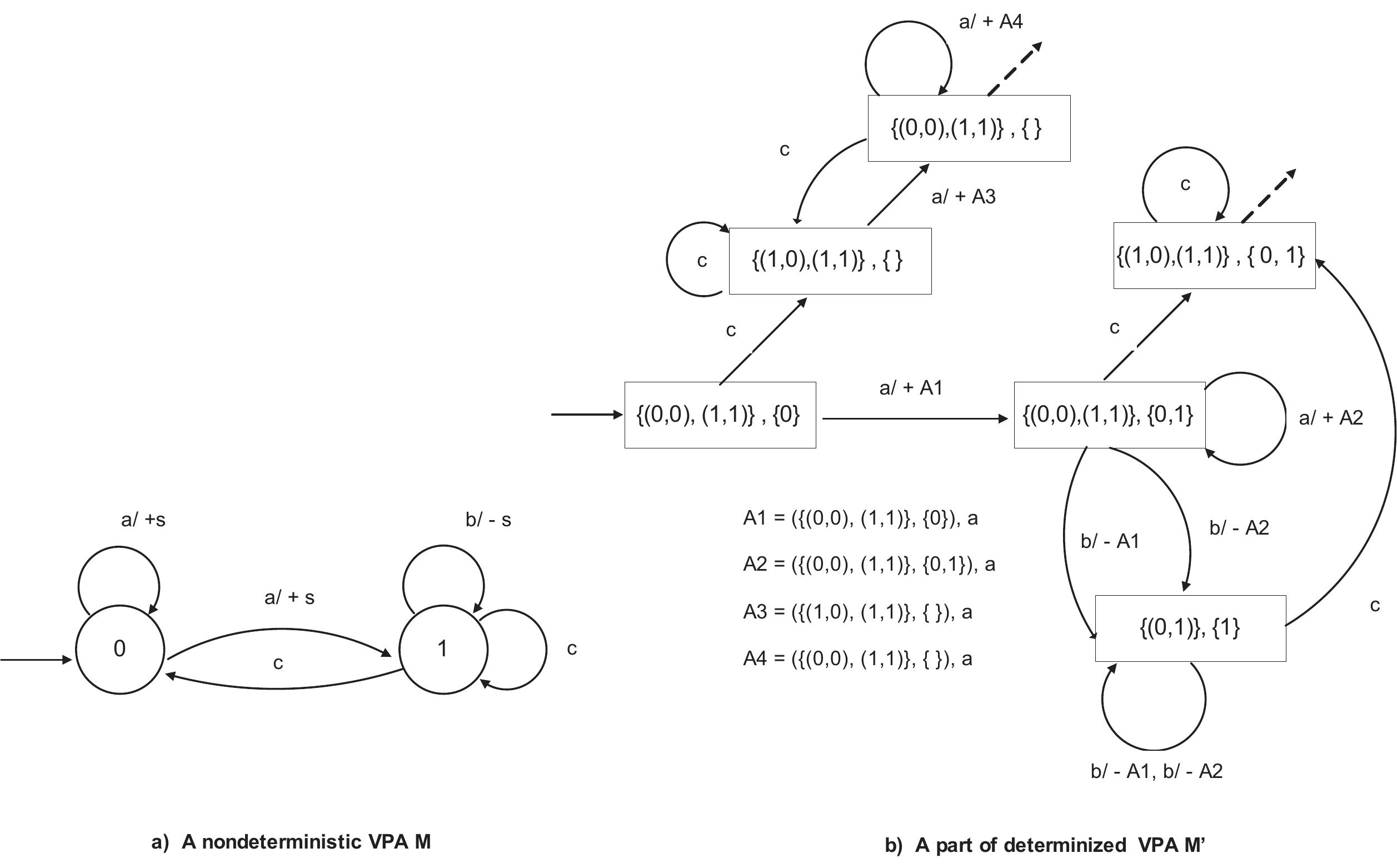}
\end{center}\caption{An example for determinization of VPA}  \label{fig:vpa}
\end{figure}

\begin{example}
We illustrate the original determinization procedure by an example
in Figure~\ref{fig:vpa}.
\end{example}

\subsection{An Improvement on Complexity for Determinization}

During implementation of VPA's operations, we found that the set of
summaries $S$ in the determinization may contain unnecessary pairs
in the sense that these pairs do not keep information of reachable
states. In other words, for any state $(S, R)$ of the determinized
VPA, $\Pi_2(S)$ does not always equal to $R$ in which $\Pi_2$ is the
projection on the second component. In the following, we present an
optimization for determinization by keeping the set of summaries as
few as possible. This simple observation, however, leads to a
tighter bound for determinization.

\subsubsection{Optimize $S$-Component}

\smallskip

We first optimize Alur-Madhusudan's determinization of VPA by
minimizing the set of summaries $S$. Given a finite set $X$, let us
denote $Id_X = \{(q,q) ~|~ q\in X\}$.

\medskip

Let $M = ( Q, \Gamma, Q_{0}, \Delta, F)$ be a nondeterministic VPA.
We construct an equivalent deterministic VPA $M^{d} = (Q', \Gamma',
Q'_{0}, \Delta',F' )$ as follows: $Q'= 2^{Q\times Q}\times 2^{Q}$,
$Q'_{0}=\{(Id_{Q_0}, Q_0)\}$ where $F' = \{(S,R) ~|~ R\cap
F\neq\varnothing\}$, $\Gamma'= Q'\times\Sigma_c$, and the transition
relation $\Delta'=\Delta'_i\cup\Delta'_c\cup\Delta'_r$ is given by:

\begin{itemize}
\item {\bf Internal:} For every $a\in\Sigma_i$,
$(S,R)\xrightarrow{a} (S',R')\in\Delta'_i$ where $S' = \{(q,q') ~|~
\exists q^{\prime\prime}\in Q: (q,q^{\prime\prime})\in S,
q''\xrightarrow{a} q'\in \Delta_i\}$ and $R' = \{q' ~|~ \exists q\in
R: q\xrightarrow{a} q'\in\Delta_i\}$

\item {\bf Push:} For every $a\in\Sigma_c$,
$(S,R)\xrightarrow{a/+(S,R,a)} (Id_{R'},R') \in\Delta'_c$ where $R'
= \{q' ~|~ \exists q\in R:
        q\xrightarrow{a/+\gamma} q'\in\Delta_c\}$

\item {\bf Pop:} For every $a\in\Sigma_r$,
\begin{itemize}
\item if the stack is empty : $(S,R)\xrightarrow{a/-\perp}(S',R')\in\Delta'_r$ where
$S' = \{(q,q')~|~\exists q^{\prime\prime}\in Q:
(q,q^{\prime\prime})\in S, q''\xrightarrow{a/-\perp}
q'\in\Delta_r\}$ and $R' = \{q' ~|~ \exists q\in R:
q\xrightarrow{a/-\perp} q'\in\Delta_r\}$.

\item otherwise:
$(S,R)\xrightarrow{a/-(S',R',a')}(S^{\prime\prime},R^{\prime\prime})\in\Delta'_r$,
where
$$\left\{
  \begin{array}{lll}
   R^{\prime\prime} & = &
    \left\{ q'
    \left|
    \begin{array}{l@{}l}
    \exists q\in R': (q,q')\in Update
    \end{array}
    \right.
    \right\} \vspace{1mm}\\

   S^{\prime\prime} & = &
   \{(q,q')~|~\exists q_3\in Q: (q,q_3)\in S', (q_3,q')\in Update\}
    \vspace{1mm}\\

   Update & = &
    \left\{  (q,q')
    \left|
    \begin{array}{l@{}l}
    \exists q_1, q_2\in Q: (q_1,q_2)\in S, \\
   q\xrightarrow{a'/+\gamma}q_1\in \Delta_c,
   q_2\xrightarrow{a/-\gamma} q'\in \Delta_r \\
    \end{array}
    \right.
    \right\} \\

  \end{array}
  \right.$$
\end{itemize}
\end{itemize}

\begin{remark}
{\it The main differences of our construction with the original one
are: (1) we initiate the initial state as $(Id_{Q_0}, Q_0)$ instead
of $(Id_Q, Q_0)$; and (2) after reading a push symbol, the automaton
will enter the state $(Id_{R'}, R')$ instead of $(Id_Q, R')$.}
\end{remark}

\begin{lemma}
For a given nondeterministic VPA $M$, let $M^{d}$ be the
deterministic VPA constructed from $M$ as above. Then, $\Pi_2(S) =
R$ for any state $(S,R)$ of $M^{d}$, where $\Pi_2$ is the projection
on the second component.
\end{lemma}

\begin{proof} Since states of $M^{d}$ are generated on-the-fly
manner, we prove the lemma by induction on the length of input
words. Let $w$ be an input word.
\begin{enumerate}
\item If $|w| = 0$, the lemma holds because $Q'_{0} = (Id_{Q_0},Q_0)$
and $\Pi_2(Id_{Q_0}) = Q_0$.

\item If $|w| = 1$, then $w = a\in \Sigma$. Consider three
cases of $a$:
\begin{itemize}
\item If $a\in \Sigma_i$: Based on the construction of transitions, we have
$(Id_{Q_0},Q_0)\xrightarrow{a} (S,R)\in\Delta'_i$ where $S =
\{(q,q') ~|~ \exists q^{\prime\prime}\in Q_0: (q,q)\in Id_{Q_0},
q\xrightarrow{a} q'\in \Delta_i\}$ and $R = \{q' ~|~ \exists q\in
Q_0: q\xrightarrow{a} q'\in\Delta_i\}$. It is easy to verify that
$\Pi_2(S) = R$.

\item If $a\in \Sigma_c$: The proof is trivial.

\item If $a\in \Sigma_r$: Since the stack now is empty, the proof is
similar to the case of internal symbols.
\end{itemize}

\item If $|w| = 2$, assume that $w = a_1a_2$. The proof is trivial
for the cases: $a_1\in\Sigma_i\cup\Sigma_c\wedge
a_2\in\Sigma_i\cup\Sigma_c$; $a_1\in\Sigma_i\wedge a_2\in\Sigma_r$.
We now check the last case: $a_1\in\Sigma_c\wedge a_2\in\Sigma_r$.
After reading $a_1$, the current state of $M^{d}$ is $(S,R)$ (with
$\Pi_2(S) = R$ by the induction assumption) and the stack content is
$(Id_{Q_0},Q_0,a_1)\bot$. On reading $a_2$, a new transition of
$M^{d}$ is created:
$(S,R)\xrightarrow{a/-(Id_{Q_0},Q_0,a_1)}(S',R')\in\Delta'_r$ where
$R' = \{ q' ~|~\exists q \in Q_0: (q,q')\in Update\}$,  $S' =
   \{(q,q')~|~\exists q\in Q: (q,q)\in Id_{Q_0}, (q,q')\in
   Update\}$, and $Update = \{  (q,q') ~|~ \exists q_1, q_2\in Q: (q_1,q_2)\in S,
   q\xrightarrow{a_1/+\gamma}q_1\in \Delta_c,
   q_2\xrightarrow{a_2/-\gamma} q'\in \Delta_r\}$.
   It is easy to see in this case that $\Pi_2(S') = R' =\Pi_2(Update)$.

\item Now, let us assume that the lemma holds with $|w|= n$. Without
loss of generality, we suppose that $w = w_1a_1w_2a_2\cdots w_k$
where in $w_1$ every call is matched by a return, but there may be
unmatched returns; $w_i$ ($i = 2\cdots k$) are well-matched words,
and $a_i$ ($i = 1\cdots k$) are calls. After reading $w$, $M^{d}$
will have its stack $(S_{k-1},R_{k-1},a_{k-1})\cdots
(S_1,R_1,a_1)\bot$ and its control state will be $(S_k,R_k)$. By the
assumption, we have $\Pi_2(S_k) = R_k$. Assume that $M^{d}$ read an
input symbol $a_k$. There are three cases of $a_k$:

\begin{itemize}
\item If $a_k\in\Sigma_i$: The automaton will go to the control state
$(S',R')$. Similar to the proof for the case $|w|=1$, we get
$\Pi_2(S') = R'$.

\item If $a_k\in\Sigma_c$: The proof is trivial.

\item If $a_k\in\Sigma_r$: The automaton changes control
state to $(S',R')$ and pops the stack symbol\\
$(S_{k-1},R_{k-1},a_{k-1})$. Namely,
$(S_k,R_k)\xrightarrow{a_k/-(S_{k-1},R_{k-1},a_{k-1})}(S',R')\in\Delta'_r$,
$$\left\{
  \begin{array}{lll}
   R' & = &
    \left\{ q'
    \left|
    \begin{array}{l@{}l}
    \exists q \in R_{k-1}: (q,q')\in Update
    \end{array}
    \right.
    \right\} \vspace{1mm}\\

   S' & = &
   \{(q,q')~|~\exists q,q_3\in Q: (q,q_3)\in S_{k-1}, (q_3,q')\in Update\}
    \vspace{1mm}\\

   Update & = &
    \left\{  (q,q')
    \left|
    \begin{array}{l@{}l}
    \exists q_1, q_2\in Q: (q_1,q_2)\in S_k, \\
   q\xrightarrow{a_1/+\gamma}q_1\in \Delta_c,
   q_2\xrightarrow{a_2/-\gamma} q'\in \Delta_r \\
    \end{array}
    \right.
    \right\} \\
  \end{array}
  \right.$$
\end{itemize}

Since $\Pi_2(S_{k-1}) = R_{k-1}$, we obtain that $\Pi_2(S') = R'$.
The lemma is proved.
\end{enumerate}
\end{proof}

\subsubsection{Remove $R$-Component}

\smallskip

As can be seen in the previous section, the component $S$ in a state
of $M^{od}$ satisfies the condition $\Pi_2(S) = R$. Therefore, we
can further optimize this determinization procedure by using the
second component of the summary $S$ as the set of reachable states.

\medskip

Let $M = (Q, \Gamma, Q_{0}, \Delta, F)$ be a nondeterministic VPA.
We construct an equivalent deterministic VPA $M^{od} = (Q', \Gamma',
Q'_{0}, \Delta',F' )$ as follows: $Q'= 2^{Q\times Q}$, $Q'_{0}=
Id_{Q_0}$ where $F' = \{S ~|~ \Pi_2(S)\cap F\neq\varnothing\}$,
$\Gamma'= Q'\times\Sigma_c$, and the transition relation
$\Delta'=\Delta'_i\cup\Delta'_c\cup\Delta'_r$ is given by:
\begin{itemize}
\item {\bf Internal:} For every $a\in\Sigma_i$,
$S\xrightarrow{a} S'\in\Delta'_i$ where $S' = \{(q,q') ~|~ \exists
q^{\prime\prime}\in Q: (q,q^{\prime\prime})\in S, q''\xrightarrow{a}
q'\in \Delta_i\}$.

\item {\bf Push:} For every $a\in\Sigma_c$,
$S\xrightarrow{a/+(S,a)} Id_{R'} \in\Delta'_c$ where $R' = \{q' ~|~
\exists q\in \Pi_2(S):
        q\xrightarrow{a/+\gamma} q'\in\Delta_c\}$.

\item {\bf Pop:} For every $a\in\Sigma_r$,
\begin{itemize}
\item if the stack is empty : $S\xrightarrow{a/-\perp}S'\in\Delta'_r$ where
$S' = \{(q,q')~|~\exists q^{\prime\prime}\in Q:
    (q,q^{\prime\prime})\in S, q''\xrightarrow{a/-\perp}
    q'\in\Delta_r\}$.
\item otherwise:
$S\xrightarrow{a/-(S',a')} S^{\prime\prime} \in\Delta'_r$, where
$$\left\{
  \begin{array}{lll}
   S^{\prime\prime} & = &
   \{(q,q')~|~\exists q_3\in Q: (q,q_3)\in S', (q_3,q')\in Update\}
    \vspace{1mm}\\

   Update & = &
    \left\{  (q,q')
    \left|
    \begin{array}{l@{}l}
    \exists q_1, q_2\in Q : (q_1,q_2)\in S, \\
   q\xrightarrow{a'/+\gamma}q_1\in \Delta_c,
   q_2\xrightarrow{a/-\gamma} q'\in \Delta_r \\
    \end{array}
    \right.
    \right\} \\

  \end{array}
  \right.$$
\end{itemize}
\end{itemize}

The next theorem immediately follows from the above construction.

\begin{theorem} For a given nondeterministic VPA $M$ of $n$ states. One can construct
a deterministic VPA $M^{od}$ such that $L(M^{od}) = L(M)$. Moreover,
the number of states and stack symbols of $M^{od}$ in the worst case
are $2^{n^2}$ and $|\Sigma_c|\cdot 2^{n^2}$, respectively.
\end{theorem}

\begin{figure}
\begin{center}
  \includegraphics[height=3in,width=5in]{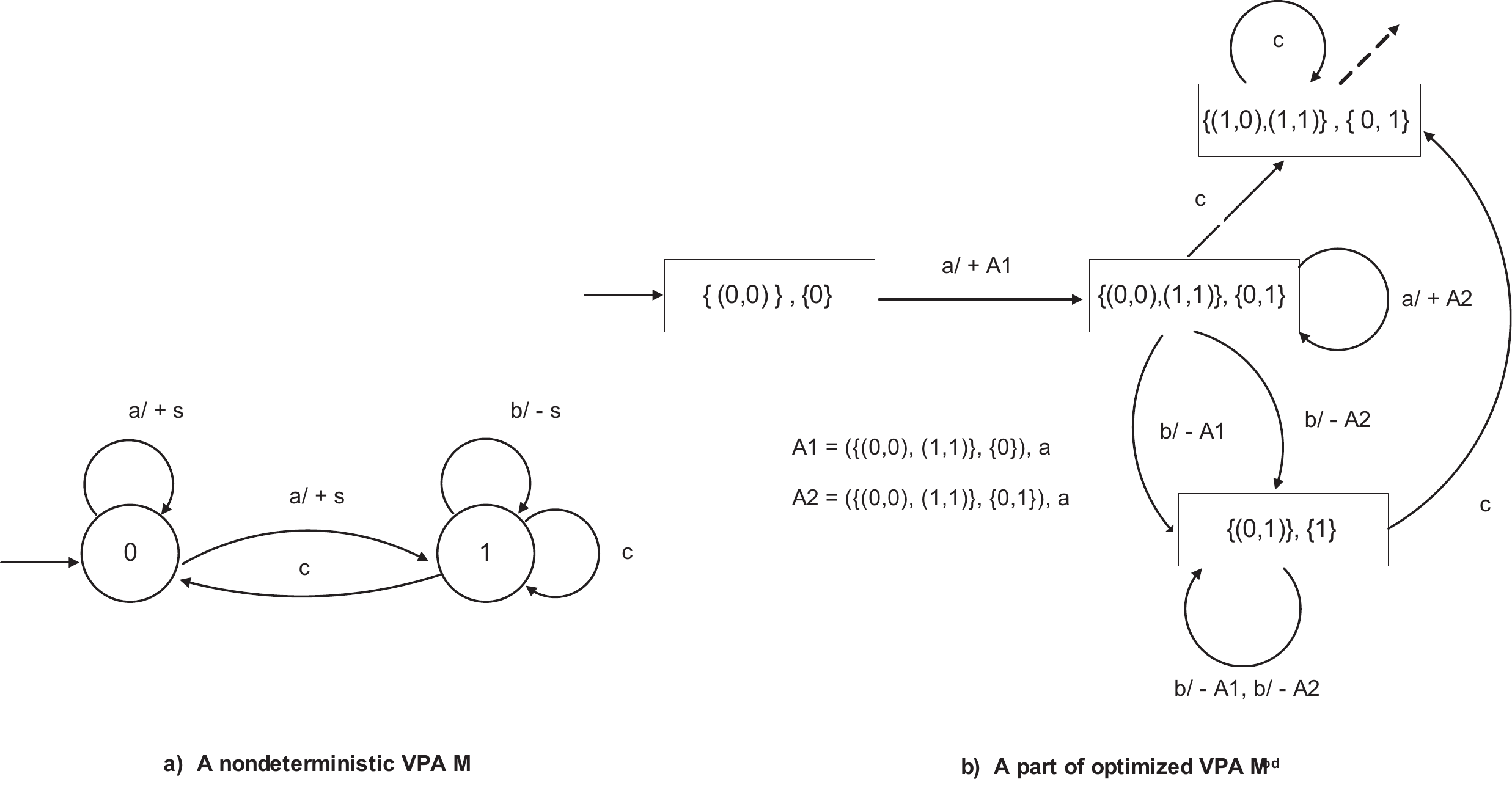}
\end{center}\caption{An example for optimized determinization of VPA}  \label{fig:vpao}
\end{figure}

\begin{example}
We illustrate the optimized procedure by determinizing non
deterministic VPA $M$ in Figure~\ref{fig:vpa}. The result of this
optimized determinization is given in Figure~\ref{fig:vpao}. We can
see that the size of the determinized VPA is reduced.
\end{example}

\begin{remark}
We should mention a fact that the model of \textit{nested words} was
proposed in~\cite{AM09} for representation of data with both a
linear ordering and a hierarchically nested matching of items.
Recall that the input word of VPA has an implicit nesting structure
defined by matching occurrences of symbols in $\Sigma_c$ with
symbols in $\Sigma_r$. In nested words, this nesting is given
explicitly, and thus they defined finite-state acceptors (with out
stacks) for nested words, so-called nested word automata. One can
interpret a \textit{nested word} automaton as a visibly pushdown
automaton over classical words. As shown in~\cite{AM09}, a
nondeterministic nested word automaton with $n$ states can be
translated into a deterministic nested word automaton with at most
$2^{n^2}$ states. In this paper, we show that the direct
determinization of VPAs can be made tighter. As stack-based
implementation is the most natural way in modeling recursive
programs, we hope that our simple improvement on determinization
procedure of VPAs is still useful.
\end{remark}

\section{Universality and Inclusion Checking}\label{Sec:Decision}

According to visibility and determinizablity, the class of VPAs is
closed under union and intersection, and complementation. Moreover,
it has been shown that the universality and inclusion problems are
\textsc{EXPTIME}-complete~\cite{AM04}.

\subsection{Emptiness Checking}

A \textit{pushdown system}(see~\cite{BEM97,Esparza00}, for example)
is pushdown automaton that is regardless of input symbols. Bouajjani
\emph{et al.}~\cite{BEM97} have introduced an efficient symbolic
method to compute reachable configurations of a pushdown system
(This method was extended for model checking LTL properties of
pushdown systems by Esparza \emph{et
al.}~\cite{Esparza00,Esparza03}). The key of their technique is to
use a finite automaton so-called \emph{$\mathcal{P}$-automaton} to
encode a set of infinite configurations of a pushdown system. It is
easy to see that the $\PP$-automaton technique can be used to solve
emptiness problem for pushdown automata (or, visibly pushdown
automata). To check the emptiness of a pushdown automaton $M$, the
first step is to compute the set of its reachable configurations
using $\PP$-automata. Second, if there exists an accepting
configurations, we conclude that the language of $M$ is not empty.

\medskip

In the following, we adapt $\PP$-automata technique to checking
emptiness of visibly pushdown automaton. Our definition, though in
essence do not differ from the one
in~\cite{BEM97,Esparza00,Esparza03}, has been tailored so that
concepts discussed in this paper are easily related to the
definition. Given a VPA $\PP = (Q, \Gamma,Q_0,\Delta, F)$, a
$\PP$-automaton is used in order to represent sets of configurations
$C$ of $\PP$. A $\PP$-automaton uses $\Gamma$ as the input alphabet,
and $Q$ as set of initial states. Formally,

\begin{definition}[$\PP$-automata]
{\it
\begin{enumerate}
\item A $\PP$-automaton of a VPA $\PP$ is a finite automaton $A =
(P,\Gamma,\delta,Q,F_A)$ where $P$ is the finite set of states,
$\delta\subseteq P\times \Gamma\times P$ is the set of transitions,
$Q$ is the set of {\em initial states} and $F_A\subseteq P$ is the
set of {\em final states}.

\item A $\PP$-automaton {\em accepts} or {\em recognizes} a
configuration $(p,w)$ if $p\xrightarrow{w} q$, for some $p\in Q$,
$q\in F_A$. The set of configurations recognized by $\PP$-automaton
$A$ is denoted by $Conf(\PP)$.
\end{enumerate}}
\end{definition}

\smallskip

For a VPA $\PP = (Q, \Gamma, Q_0,\Delta, F)$ and the set of
configurations $C$, let $A$ be a $\PP$-automaton representing $C$.
The $\PP$-automaton $A_{post^*}$ representing the set of
configurations reachable from $C$ ($Post^*(C)$) is constructed as
follows: We compute $Post^*(C)$ as a language accepted by a
$\PP$-automaton $A_{post^*}$ with $\epsilon$-moves. We denote the
relation  $q (\xrightarrow{\epsilon})^*
\cdot\xrightarrow{\gamma}\cdot(\xrightarrow{\epsilon})^* \cdot p$ by
$\Longrightarrow^{\gamma}$. Formally, $A_{post^*}$ is obtained from
$A$ in two stages:
\begin{itemize}
\item For each pair $(q',\gamma')$ such that $\PP$ contains at least one rule of the form
$q\xrightarrow{\text{$a/+\gamma'$}} q' \in\Delta_c$, add a new state
$p_{(q',\gamma')}$ to $A$.

\item Add new transitions to $A$ according to the following
saturation rules:
\end{itemize}

\begin{algorithm}
\begin{enumerate}

\item \textbf{Internal:} If $q \xrightarrow{a} q'\in\Delta_i$ and
$q\Longrightarrow^{\gamma} p$ in the current automaton, add a
transition $(q',\gamma,p)$.


\item \textbf{Push:} If $q \xrightarrow{\text{$a/+\gamma'$}} q'\in\Delta_c$
and $q\Longrightarrow^{\gamma} p$ in the current automaton, first
add $(q',\gamma',p_{(q',\gamma')})$, and then add
$(p_{(q',\gamma')}, \gamma,p)$.

\item \textbf{Pop:} If $q \xrightarrow{\text{$a/-\gamma$}} q'\in\Delta_r$ and
$q\Longrightarrow^{\gamma} p$ in the current automaton, add a
transition $(q',\epsilon,p)$.

\end{enumerate}
\end{algorithm}

\begin{figure}
\begin{center}
  \includegraphics[height=2.4in,width=5in]{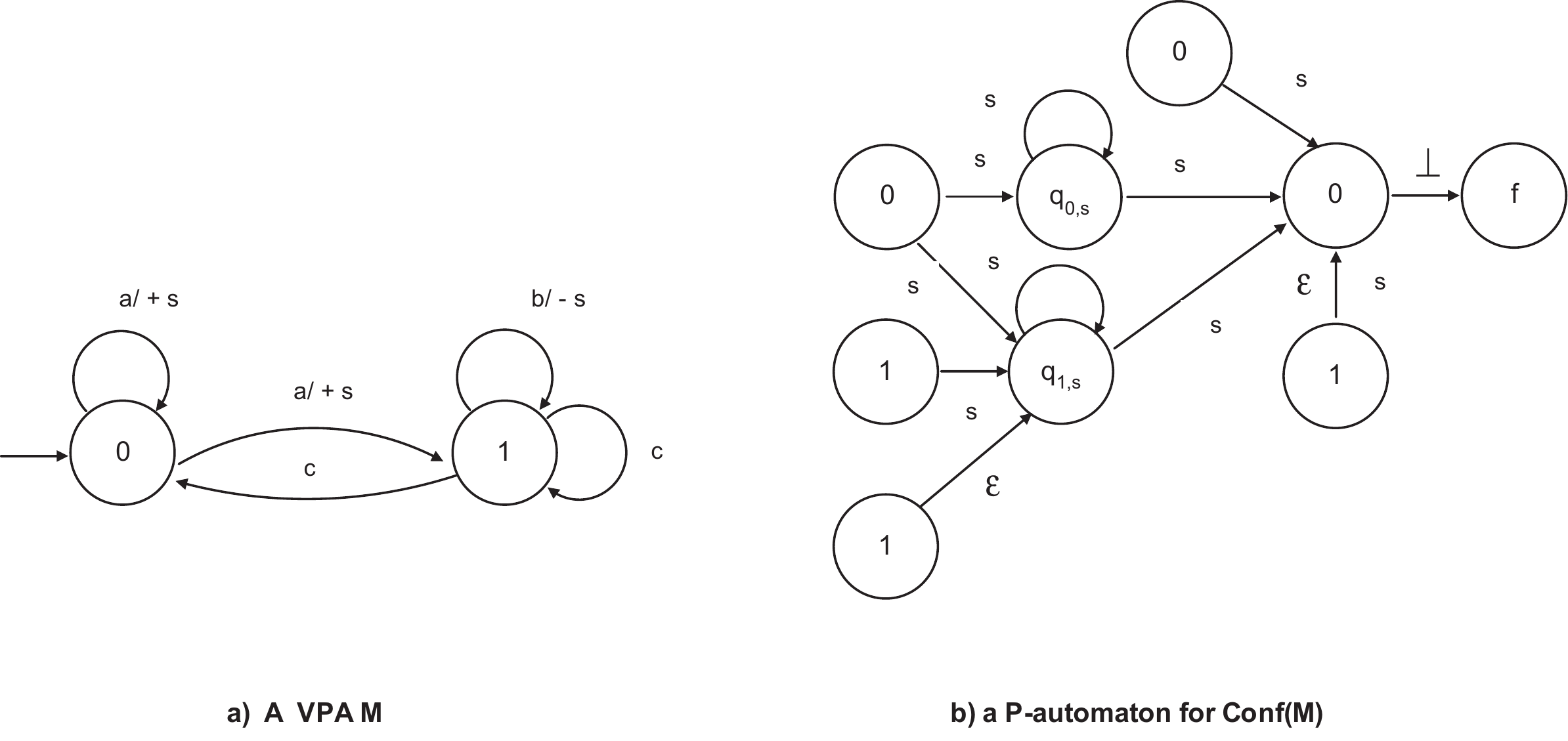}
\end{center}\caption{An example of $\PP$-automata}  \label{fig:PA}
\end{figure}

\begin{example}
Let us revisit nondeterministic VPA $M$ in Figure~\ref{fig:vpa}. A
$\PP$-automaton for the set of all reachable configurations of $M$
is given in Figure~\ref{fig:PA}.
\end{example}

\subsection{Universality Checking}

\smallskip

In this section, we propose an \EMPH{on-the-fly} method to solve the
universality and inclusion problems for visibly pushdown automata.
We first briefly recall the standard method in the next subsection.

\subsubsection{Standard Methods}

\smallskip

The standard algorithm for universality of VPA is to first
determinize the automaton, and then check for the reachability of a
non-accepting states. Reachable configurations of a determinized VPA
can be computed by using $\PP$-automata technique. A configuration
$c = (q,w)$ is said a rejecting configuration if $q$ is not a final
location. Whenever a rejecting configuration is found, we stop and
report that the original VPA is not universal. Otherwise, if all
reachable configurations of determinized VPA are accepting
configurations, the original VPA is universal.

\subsubsection{On-the-fly Methods}

\smallskip

To improve efficiency of checking, we perform simultaneously
on-the-fly determinization and $\PP$-automata construction. There
are two interleaving phases in this approach. First, we determinize
VPA $M$ step by step (iterations). After each step of
determinization, we update the $\PP$-automaton. Then, using the
$\PP$-automaton, we perform determinization again, and so on. It is
crucial to note that this procedure terminates. This is because the
size of the $M^{od}$ is finite, and the $\PP$-automaton construction
is terminated. However, once a rejecting state is added to the
$\PP$-automaton, we stop and report that the VPA is not universal.
Let $\textsf{Conf}(M^{od})$ and $\textsf{Rejecting-Conf}(M^{od})$
denote the sets of reachable and rejecting configurations of
$M^{od}$, respectively. With the above observation, the following
lemma holds:

\begin{lemma}\label{lem:key}
Let $M$ be a nondeterministic VPA. The automaton $M$ is not
universal iff there exists a rejecting reachable configuration of
$M^{od}$, \emph{i.e.,} {\em
$\textsf{Conf}(M^{od})\cap\textsf{Rejecting-Conf}(M^{od})\neq
\varnothing$}.
\end{lemma}

Therefore checking universality of $M$ amounts to finding a
rejecting configuration of $M^{od}$. In Algorithm~\ref{Al:Onthefly},
we present an on-the-fly way to explore such rejecting
configurations.

\begin{algorithm}[h]
\caption{On-the-fly algorithm} \label{Al:Onthefly}
\begin{algorithmic}

\STATE {\textbf{Input:} A nondeterministic VPA $M = (Q, Q_0,\Gamma,
\Delta, F)$} \STATE {\textbf{Result:} Universality of $M$}

\medskip

     \STATE {\textbf{begin}}
          \STATE {Create the initial state of the determinized VPA $M^{od}$;}
          \STATE {Initiate $\PP$-automaton $A$ to present the initial configuration of $M^{od}$;}
          \STATE {$A_{post^*} \longleftarrow A$;}
          \STATE {Create transitions of $M^{od}$ departing from the initial state;}

\medskip

    \WHILE {(the set of new transitions of $M^{od}$ is not empty)}
        \STATE {Update the $\PP$-automaton $A_{post^*}$ using new transitions of
    $M^{od}$;}
        \IF{a rejecting state is added to $A_{post^*}$}
              \RETURN{ False;}
        \ENDIF

        \STATE{Update $M^{od}$ using new transitions of $A_{post^*}$;}
    \ENDWHILE

    \RETURN{True;}
\STATE{\textbf{end}}
\end{algorithmic}
\end{algorithm}

Having said this, time complexity of the on-the-fly method is same
as the complexity of the standard one. However, if the input VPA is
not universal, the on-the-fly method is significantly faster. This
is because the on-the-fly method does not need to perform full
determinization, and thus it will immediately stop whenever a
rejecting state is found.

\begin{figure}
\begin{center}
  \includegraphics[height=3.5in,width=5in]{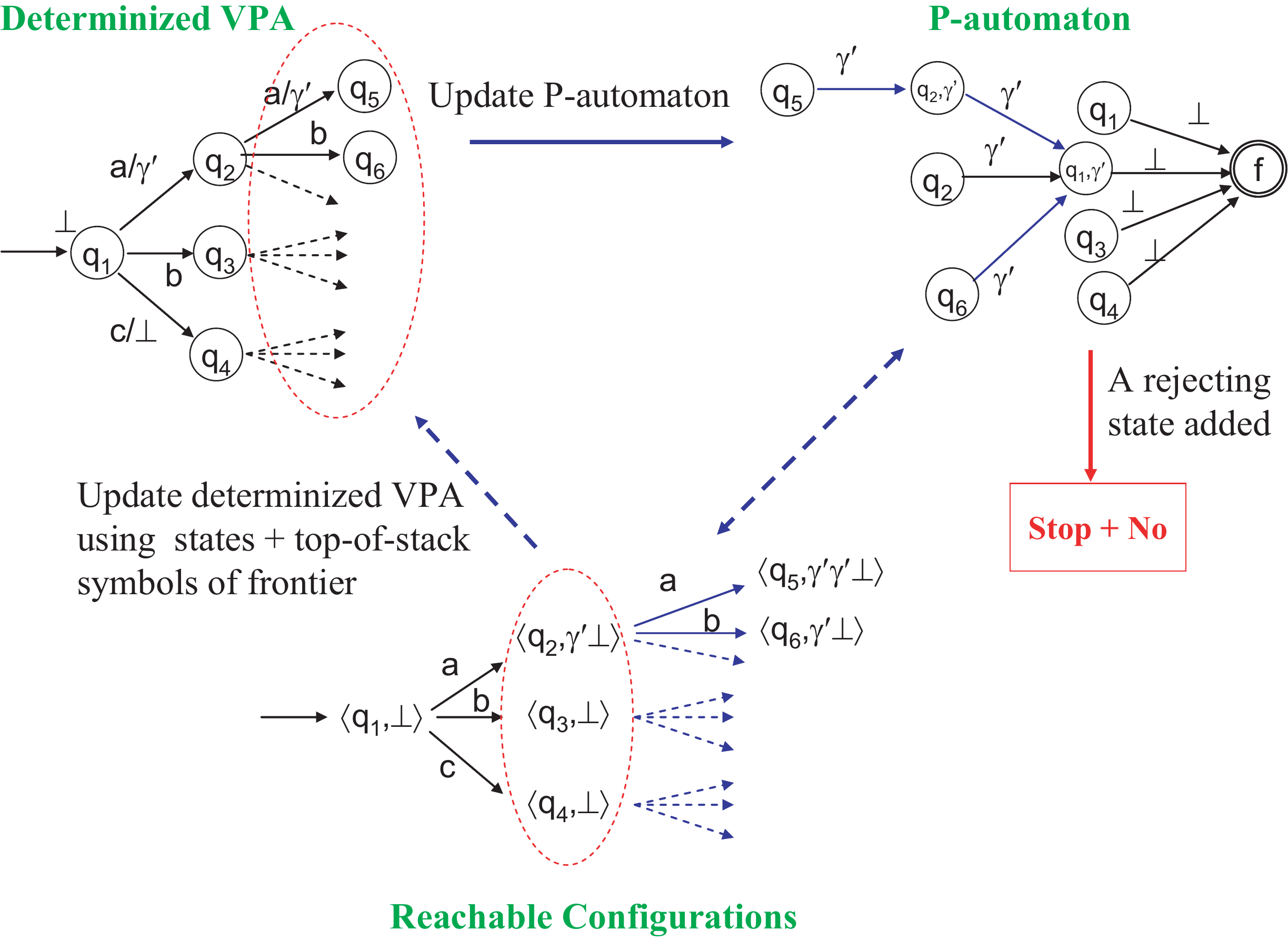}
\end{center}\caption{Simulation for On-the-fly
  Method}  \label{fig:onthefly-process}
\end{figure}

\begin{example}\label{Ex:on-the-fly}
{\it We illustrate the on-thy-fly algorithm by an example given in
Figure~\ref{fig:onthefly-process}. We assume that $a\in\Sigma_c$,
$b\in\Sigma_i$, and $c\in\Sigma_r$. The process of the algorithms is
performed as below:

\begin{enumerate}
\item At the first time, assume that the initial state $q_1$ of determinized VPA
$M^{od}$ is created.

\item Then, the $\PP$-automaton $A$ is constructed which includes two
states $\{q_1, f\}$ and one transition
$q_1\xrightarrow{\text{$\bot$}} f$, where $f$ is a unique final
state. $\PP$-automaton $A$ represents a set of initial
configurations $\{(q_1,\bot)\}$ of $M^{od}$.

\item Update $M^{od}$ using $A$. Suppose that $M^{od}$ has new states
$\{q_2,q_3,q_4\}$; and new transitions
$\{q_1\xrightarrow{\text{$a/+\gamma'$}} q_2, q_1\xrightarrow{b} q_3,
q_1\xrightarrow{\text{$c/-\bot$}} q_4\}$.

\item Update $\PP$-automaton $A$ using new transitions of $M^{od}$. $A$
has new states $\{q_2,q_3,q_4,p_{(q_1,\gamma')}\}$ and transitions
$\{ q_2\xrightarrow{\text{$\gamma'$}} p_{(q_1,\gamma')},
p_{(q_1,\gamma')}\xrightarrow{\text{$\bot$}} f,
q_3\xrightarrow{\text{$\bot$}} f, q_4\xrightarrow{\text{$\bot$}}
f\}$.

\item Again, update $M^{od}$ using new transitions of $A$,
and so on.
\end{enumerate}}
\end{example}

\subsection{Inclusion Checking}

\smallskip

Let $A$ and $B$ be two VPAs. We want to check whether $L(A)\subseteq
L(B)$. The standard method is to check whether
$L(A\times\overline{B}) =\varnothing$, where $\overline{B}$ is the
complement of $B$.

\medskip

The on-the-fly approach tries to find if there exists at least a
word $w\in L(A)\setminus L(B)$. If such a word $w$ was found, we can
conclude that $L(A)\nsubseteq L(B)$. Otherwise, $L(A)$ is a subset
of $L(B)$. To do so, similar to the case of universality checking,
we perform on-the-fly determinization for $B$ and simultaneously
$\PP$-automata construction for the product VPA $A\times B^{od}$,
where $B^{od}$ is determinized counterpart of $B$. Once a state
$(p,q)\in (F_A \times (Q_{B^{od}}\backslash F_{B^{od}})$ is added to
the $\PP$-automaton. There exists a word $w$ such that, after
reading $w$, $A$ leads to an accepting configuration whereas
$B^{od}$ leads to a rejecting configuration. This means that there
exists a word $w\in L(A)\setminus L(B)$. In this case, we stop and
report that $L(A)\nsubseteq L(B)$.

\medskip

It is crucial to note that, if $L(A)\subseteq L(B)$, the on-the-fly
approach needs to fully determinize $B$, and this is similar to the
standard approach. Therefore, in the worst case, the time complexity
of the on-the-fly approach equals to that of the standard one.

\section{Implementation and Experiments}\label{Sec:Implement}

We have implemented the above approaches for testing universality
and inclusion of VPA in a prototype tool. The package is implemented
in Java 1.5.0 on Windows XP. To compare the on-the-fly algorithm
with the standard algorithm, we run our implementations on randomly
generated VPAs. All tests are performed on a PC equipped with 1.50
GHz Intel\textregistered\ Core\texttrademark\ Duo Processor L2300
and 1.5 GB of memory.

\medskip

During experiments, we fix the size of the input alphabet to
$|\Sigma_c|=|\Sigma_r| =|\Sigma_i| = 2$, and the size of the stack
alphabet to $|\Gamma| = 3$. We first set parameters of the tests as
follows:
\begin{definition}[random 1]\label{Def:Random1}
The \emph{density of final states} $f =\frac{|F|}{|Q|}=1$ and the
\emph{density of transitions} $r = \frac{k_a}{|Q|}=2$, where $k_a$
is the number of transitions for each input symbol $a$.
\end{definition}

We ran our tests on randomly VPA generated by the parameter random
1. We have tried VPAs sizes from 10 to 100. We generated 50 VPAs for
each sample point, and setting timeout to 60 seconds. The
experimental results are given in Table~\ref{table:random1}. We
found that all successfully checked VPAs are not universal, and thus
we omit the row for universal results in the table. The experiments
shows that \textsf{STANDARD} can solve for generated VPA instances
with 5 states only. It gets stuck when the number of states greater
than or equal to 10. Meanwhile, \textsf{ON-THE-FLY} is significantly
efficient than \textsf{STANDARD}, they can check for almost VPAs.

\begin{table}
\begin{center}
\caption{Universality checking for VPA generated by random 1}
\label{table:random1}
\begin{tabular}{@{}l@{\quad}rrr@{~~}rrr@{~~}rrr@{~~}rr@{}}
\hline\hline & \multicolumn{11}{c}{\textit{number of states}} \\
\textsf{ON-THE-FLY} & $\mathtt{5}$ & $\mathtt{10}$ & $\mathtt{20}$ &
$\mathtt{30}$ & $\mathtt{40}$ & $\mathtt{50}$ & $\mathtt{60}$ &
$\mathtt{70}$ & $\mathtt{80}$ & $\mathtt{90}$ & $\mathtt{100}$ \\
\hline
success
& 50 & 50 & 50 & 50 & 50 & 50 & 50 & 50 & 50 & 50 & 46  \\
total time & 23 & 46 & 52 & 71 & 110 & 186 & 210 & 274  & 247 & 407
& 686
\\[.5ex]
timeout number (\emph{60} s) & 0 & 0 & 0 & 0 & 0 & 0 & 0 & 0 & 0 & 0
& 4 \\
\hline & \multicolumn{11}{c}{\textit{number of states}} \\
\textsf{STANDARD} & $\mathtt{5}$ & $\mathtt{10}$ & $\mathtt{20}$ &
$\mathtt{30}$ & $\mathtt{40}$ & $\mathtt{50}$ & $\mathtt{60}$ &
$\mathtt{70}$ & $\mathtt{80}$ & $\mathtt{90}$ & $\mathtt{100}$ \\
\hline
success
& 21 & 1 & 0 & 0 & 0 & 0 & 0 & 0 & 0 & 0 & 0 \\
total time & 456 & 31 & 0 & 0 & 0 & 0 & 0 & 0 & 0 & 0 & 0 \\[.5ex]
timeout number (\emph{60} s) & 29 & 49 & 50 & 50 & 50 & 50 & 50 & 50
& 50 &
50 & 50 \\
\hline
\end{tabular}
\end{center}
\end{table}

The parameter random 1 does not guarantee the completeness of VPAs.
Therefore, the probability of being universal is very low. In order
to increase the probability of being universal, we set a new
parameter as below:

\begin{definition}[random 2]
The \emph{density of final states}$f =\frac{|F|}{|Q|}$ and the
\emph{density of transitions} $r: Q\times\Sigma \rightarrow N$;
$r(q,a)$ depends on not only the input symbol $a$ but also on the
state $q$. In particular, we select $r(q,a) = 2$ for all $q\in Q$
and $a\in\Sigma_c$, $r(q,b) = 6$ for all $q\in Q$ and
$b\in\Sigma_r$, and $r(q,c) = 2$ for all $q\in Q$ and
$c\in\Sigma_i$.
\end{definition}

As can be seen, with random 2, a VPA with 10 states has 200
transitions. We again test for various sizes of VPAs from 5 to 50.
We ran with 50 samples for each point, setting timeout to 180
seconds. The results are reported in Table~\ref{table:random2}. For
this parameter, results of \textsf{STANDARD} are almost timeout even
with only 5 states. \textsf{ON-THE-FLY} behaves in significantly
better ways than those of \textsf{STANDARD}.

\begin{table}
\begin{center}
\caption{Universality checking for VPA generated by random 2, f =
0.6} \label{table:random2}
\begin{tabular}{@{}l@{\quad}rrr@{~~}rrr@{~~}rrr@{}}
\hline\hline & \multicolumn{7}{c}{\textit{number of states}} \\
\textsf{ON-THE-FLY} & $\mathtt{5}$ & $\mathtt{10}$ & $\mathtt{15}$ &
$\mathtt{20}$ & $\mathtt{30}$ & $\mathtt{40}$ & $\mathtt{50}$ \\
\hline
success
& 50 & 43 & 33 & 18 & 6 & 2 & 0 \\
total time & 68 & 1425 & 2310 & 1950 & 1024 & 345 & 0 \\[.5ex] timeout number
(\emph{180} s) & 0 & 7 & 17 & 32 & 44 & 48 & 50 \\
\hline & \multicolumn{7}{c}{\textit{number of states}} \\
\textsf{STANDARD} & $\mathtt{5}$ & $\mathtt{10}$ & $\mathtt{15}$ &
$\mathtt{20}$ & $\mathtt{30}$ & $\mathtt{40}$ & $\mathtt{50}$ \\
\hline
success
& 20 & 0 & 0 & 0 & 0 & 0 & 0 \\
total time & 3240 & 0 & 0 & 0 & 0 & 0 & 0 \\[.5ex]
timeout number (\emph{180} s) & 0 & 50 & 50 & 50 & 50 & 50 & 50 \\
\hline
\end{tabular}
\end{center}
\end{table}

\medskip

We also performed experiments for inclusion checking $L(A)\subseteq
L(B)$. For this, we selected parameter random 2 for $f = 0.5$. We
generated various sizes of $A$ (10, 100, 200, 500, 1000, and 3000
states) and $B$ (5 and 10 states). We ran with 20 samples for each
point, setting timeout to 300 seconds. For this test,
\textsf{STANDARD} does not work well, it get all timeout for the
smallest size $(10,5)$. Meanwhile, \textsf{ON-THE-FLY} behaves in a
significant way. The detailed experimental results of
\textsf{ON-THE-FLY} for inclusion checking are reported in
Table~\ref{table:inclusionregular}.

\begin{table}
\begin{center}
\caption{Checking inclusion with $r(q,a) = 2$, $f = 0.5$}
\label{table:inclusionregular}
\begin{tabular}{@{}l@{\quad}ccc@{~~~~}ccc@{}}
\hline\hline & \multicolumn{6}{c}{\textit{number states of A and B}}\\
\textsf{ON-THE-FLY} & (10,5) & (100,5) & (200,5) & (500,5) & (1000,5) & (3000,5)\\
\hline success
& 20 & 20  & 15 & 7 & 5 & 2\\
total time & 27 & 910 & 830 & 336 & 1257 & 357 \\[.5ex]
timeout number (\emph{300} s) & 0 & 0 & 13 & 15 & 15 & 18\\
\hline
\end{tabular}
\end{center}
\end{table}

\section{Related Work}\label{Sec:Related}

The model of \textit{nested words} was proposed in~\cite{AM09} for
representation of data with both a linear ordering and a
hierarchically nested matching of items. Recall that the input word
of VPA has an implicit nesting structure defined by matching
occurrences of symbols in $\Sigma_c$ with symbols in $\Sigma_r$. In
nested words, this nesting is given explicitly, and thus they
defined finite-state acceptors (with out stacks) for nested words,
so-called nested word automata. One can interpret a \textit{nested
word} automaton as a visibly pushdown automaton over classical
words. As shown in~\cite{AM09}, a nondeterministic nested word
automaton with $n$ states can be translated into a deterministic
nested word automaton with at most $2^{n^2}$ states. In this paper,
we show that the direct determinization of VPAs can be made tighter.
As stack-based implementation is the most natural way in modeling
recursive programs, we hope that our simple improvement on
determinization procedure of VPAs is still useful.

\medskip

The first implementation of VPA, named
VPAlib~\footnote{http://www.emn.fr/x-info/hnguyen/vpa/}, only works
for basic operations such as union, intersection, and
determinization. In their implementation, however, determinization
was performed in an exhaustive way. Namely, unreachable states and
redundant transitions were also generated. Therefore their
determinization easily gets stuck with VPAs of small size. We
implemented our prototype tool upon the top of \textsf{VPAlib}. In
particular, we first reused and improved data structures as well as
basic operations of \textsf{VPAlib}. Next, we implemented
determinization on-the-fly manner, in which only reachable states
and necessary transitions were created. Then, we used $\PP$-automata
technique to check emptiness (as well as computing reachable
configurations) of VPAs. Finally, we implemented the standard and
on-the-fly methods to check universality and inclusion of VPAs.

\section{Conclusion}\label{Sec:Conclusion}

In this paper we have shown that the upper bound for determinization
of VPA can be made tighter. Our improvement comes from a simple
observation that, in Alur-Madhusudan determinization procedure, the
set of summaries S may contain unnecessary pairs in the sense that
these pairs do not keep information of reachable states. We exploit
this observation to present a new algorithm for determinization by
keeping the second component of S always equal to R. This leads to
an optimization of the determinization algorithm by using the second
component of the summary edge S as the set of reachable states R and
this permits to construct a deterministic $VPA$ with only $2^{n ^2}$
states.

\medskip

We also have presented on-the-fly algorithms for testing
universality and inclusion of nondeterministic VPAs. In summary, to
check universality of a nondeterministic VPA $M$, the intuition
behind on-the-fly manner is try to find whether there exists a word
$w$ such that $w \notin L(M)$. Similarly, to check inclusion $L(M)
\subseteq L(N)$, the ideas behind is to find whether there exists at
least a word $w$ such that $w \in L(M)\backslash L(N)$. All
algorithms has been implemented in a prototype tool. Although the
ideas of the on-the-fly methods are simple, the experimental results
showed that the proposed algorithms are considerably faster than the
standard ones, especially for the cases universality / inclusion do
not hold.

\medskip

Finally, we should emphasize that we need to improve our tool (as
well as algorithms) to check larger examples. On the other hand, we
also need to consider to apply the tool to case studies in practice.
At the moment, the data structures for VPA are rather naive. That is
why the running time of our tool is not fast. It would be
interesting to explore a more compact data structure. For this, we
plan to manipulate VPA using BDD-based representation. Despite these
many limitations, however, we believe that this paper provides a
first stepping-stone for developing a VPA-based model checker.

\medskip

\textbf{Acknowledgements:} I would like to thank Professor Mizuhito
Ogawa and Nao Hirokawa for helpful discussions on this work. Thanks
also go to Ha Nguyen for fruitful discussions about
\textsf{VPAlib}'s source code, and anonymous referees for their
valuable comments and suggestions in improving the paper. Last but
definitely not least, I am grateful to Axel Legay for his comments
and support in preparing this final version.

\medskip

\bibliographystyle{eptcs}

\end{document}